\documentclass{aa}
\usepackage{graphicx}
\usepackage{txfonts}
\usepackage{hyperref}
\usepackage{blindtext}
\usepackage{multirow}
\hypersetup{colorlinks=true,linkcolor=blue,citecolor=blue,filecolor=blue,urlcolor=blue,}

\newcommand{\kms} {\TextOrMath{\,km s$^{-1}$}{\mathrm{\,km s}^{-1}}}
\newcommand{\msol}{\TextOrMath{\,M$_\odot$}{\mathrm{\,M}_\odot}} 
\newcommand{\lsol} {\TextOrMath{\,L$_\odot$}{\mathrm{\,L}_\odot}}

\newcommand{\bpass}{\texttt{BPASS}v2.2}

\begin{document} 

\title{Binary neutron star merger offsets from their host galaxies}
\subtitle{GW~170817 as a case study}

\author{
N. Gaspari\inst{1}
\and
H.~F. Stevance\inst{2,3}
\and
A.~J. Levan\inst{1,4}
\and
A.~A. Chrimes\inst{5,1}
\and
J.~D. Lyman\inst{4}
}

\institute{
Department of Astrophysics/IMAPP, Radboud University, P.O. Box 9010, 6500 GL Nijmegen, the Netherlands\\
\email{nicola.gaspari@live.it}
\and
Astrophysics sub-Department, Department of Physics, University of Oxford, Keble Road, Oxford, OX1 3RH, UK
\and
Astrophysics Research Centre, School of Mathematics and Physics, Queen’s University Belfast, BT7 1NN, UK
\and  
Department of Physics, University of Warwick, Coventry CV4 7AL, UK
\and 
European Space Agency (ESA), European Space Research and Technology Centre (ESTEC), Keplerlaan 1, 2201 AZ Noordwijk, the Netherlands
}

\date{Received ---; accepted ---}
 
\abstract
{}
{The locations of binary neutron star (BNS) mergers within their host galaxies encode the systemic kicks that these systems received in the supernova aftermath. We investigate how the galactic potential and the systemic kicks shape the offset distribution of BNS mergers with a case study of GW~170817 and its host NGC~4993.}
{We derived dynamical constraints on the host potential from integral field spectroscopy with Jeans anisotropic modelling. We evolved the trajectories of synthetic BNSs from the BPASS code in the galactic potential, using two different kick prescriptions to investigate how the observed offsets might differentiate between these two possibilities. The simulation was repeated after swapping the host potential with that of a dwarf galaxy, to test the effect of the potential on the offsets.}
{The location of GW~170817 is entirely consistent with our predictions, regardless of large or small kicks, because the strong potential of NGC~4993 is only diagnostic of very large kicks. In galaxies of similar or greater mass, large offsets can constrain large kicks, while small offsets do not provide much information. In an old dwarf galaxy, on the other hand, small offsets can constrain small kicks, while large offsets would prevent host association.}
{}


\maketitle


\section{Introduction}

Mergers of binary neutron stars (BNSs) were proposed to be the origin of short-duration gamma-ray bursts \cite[SGRBs,][]{1989Natur.340..126E,1992ApJ...395L..83N,1993ApJ...413L.101K} not long after the first BNS was discovered \citep{1975ApJ...195L..51H}. As of today, we have almost three decades of literature supporting BNS mergers being the main source of SGRBs through either direct or indirect evidence \cite[see for a review][]{2007NJPh....9...17L,2007PhR...442..166N,2014ARAA..52...43B,abbott2017}, although we also have evidence that SGRBs are not an unbiased sample, since it is unlikely that all BNS mergers produce an SGRB \citep{2022Natur.612..223R,2022Natur.612..228T,2022PhRvD.105h3004S,2022AA...666A.174S,2024Natur.626..737L,2024Natur.626..742Y} and that all SGRBs are produced by a BNS merger \citep{1995MNRAS.275..255T,1998ApJ...494L..57Q,2006MNRAS.368L...1L,2008MNRAS.385.1455M,2008MNRAS.385L..10T,2020ApJ...895...58G}.

One key piece of evidence that links the two phenomena is the locations of SGRBs within their host galaxies. Indeed, the distributions of SGRB offsets from the centres of their hosts do not trace the stellar light, which is both consistent with the merger times and systemic kicks expected for BNSs, and effective in restricting alternative progenitors \citep{pzy1998,bloom1999,fryer1999,bloom2002,perna2002,vosstauris2003,belczynski2006,fong2010,berger2011,fong2013,mandhai2022,fong2022,nugent2022,2024MNRAS.527.1101G}. 
The study and modelling of observed SGRBs offsets has therefore been used to inform several aspects of BNS mergers and their stellar progenitors \citep{zemp2009,berger2010,kelley2010,church2011,behroozi2014,tunnicliffe2014,abbott2017_host,zevin2020,perna2022,oconnor2022}.

To date, GW~170817/GRB~170817A represents the most robust and well-studied case of a BNS merger \citep[][and references therein]{abbott2017,2021ARAA..59..155M}, and its location together with the properties of the host galaxy NGC~4993 have been discussed in detail in several works \citep{blanchard2017,abbott2017_host,im2017,levan2017,pan2017,palmese2017,artale2019,ebrova2020,perna2022,kilpatrick2022,stevance2023}.
In this work, we expand and improve the characterisation of both the location of GW~170817, and its host galaxy NGC~4993. We derive dynamical constraints on the potential of the host directly from integral field spectroscopy (IFS) with Jeans anisotropic modelling \citep[JAM, ][]{cappellari08,cappellari20}, which allows for a better estimate of the potential than from scaling relations alone \citep[c.f.][]{church2011,abbott2017_host,zevin2020}. We combine the galactic potential with a synthetic population of BNSs to model their trajectories within NGC~4993, and then compare the predicted merger locations with that of GW~170817. Our kinematic model is meant to complement the IFS-based analysis of host galaxies and BNS progenitors of \cite{stevance2023}, by providing further constraints on the BNS merger progenitors.

The paper is structured as follows. In Sect.~\ref{sec:2}, we briefly describe the data we use. In Sect.~\ref{sec:3}, we describe the JAM method to model the galactic kinematics, and the simulations of BNS trajectories within NGC~4993. In Sect.~\ref{sec:4}, we present and analyse the results, including a comparison with a dwarf galaxy, before we summarise and conclude in Sect.~\ref{sec:5}.
Throughout the paper, magnitudes are reported in the AB system \citep{1982PASP...94..586O}. We adopt a $\Lambda$CDM cosmology with $H_0=67.66$ km s$^{-1}$ Mpc$^{-1}$, $\Omega_\Lambda=0.69$, and $\Omega_\text{m}=0.31$ \citep{2020A&A...641A...6P}.


\section{Data}\label{sec:2}

The main subject of this work is NGC~4993, the host galaxy of the BNS merger GW~170817 \citep{abbott2017,blanchard2017,levan2017}. NGC~4993 is a nearby S0 galaxy \citep{1991rc3..book.....D,2015AA...581A..10C} with a type-II shell system \citep{1983ApJ...274..534M,blanchard2017,levan2017,palmese2017,ebrova2020,kilpatrick2022}. \cite{2017ApJ...848L..31H} place the galaxy at a luminosity distance of $D_\mathrm{L}=41.0\pm 3.1$ Mpc by averaging the estimate from its Hubble-flow velocity with the estimate from the fundamental plane relation. This distance is consistent with the measures obtained from the GW signal by \cite{abbott2017}, from the fundamental plane relation by \cite{im2017}, and from surface brightness fluctuations by \cite{2018ApJ...854L..31C}, and it implies a redshift corrected for proper motions of $z=0.0092\pm 0.0007$. The optical counterpart of GW~170817 was localised at an angular offset of $10.315\pm 0.007''$ from the galaxy centroid \citep{blanchard2017,levan2017}, which at this redshift is equal to a proper distance of $2.01\pm 0.15$ kpc.

To model NGC~4993, we made use of both imaging and spectroscopy. For the photometry, we used observations obtained in the F606W filter with the camera WFC3/UVIS on board the \textit{Hubble} Space Telescope (HST) by Program GO-14771 (PI Tanvir), reduced by \cite{2017ApJ...848L..27T}. For the spectroscopy, we used the observations obtained with the integral-field spectograph MUSE on the Very Large Telescope (VLT) by Program 099.D-0668 (PI Levan), reduced by \cite{levan2017}.

For comparison, we also modelled NGC~1396, a dwarf elliptical galaxy \citep{1989AJ.....98..367F,1991rc3..book.....D} at a distance of $20.0\pm1.4$ Mpc in the Fornax cluster \citep{2009ApJ...694..556B}. For the photometry, we used the observations in F606W obtained with the camera ACS/WFC on board HST by Program GO-10129 (PI Puzia), reduced by \cite{2014ApJ...786...78P}. For the spectroscopy, we used the observation obtained with MUSE at the VLT by Program 094.B-0895 (PI Lisker), reduced by \cite{2016MNRAS.463.2819M}.


\section{Methods}\label{sec:3}

\subsection{Host galaxy modelling}

\subsubsection{Surface brightness}
To model the surface brightness, $\Sigma(x',y')$, we used the multi-Gaussian expansion (MGE) formalism \citep{emsellem94,cappellari02}, in which $\Sigma$ is expressed as a sum of $N$ Gaussian functions,
\begin{equation}\label{eq:1}
\Sigma(x',y')=\sum^{N}_{n=1}\frac{L_n}{2\pi\sigma_n^2q'_n}\exp\left[ -\frac{1}{2\sigma_n^2} \left( {x'}^2 +\frac{{y'}^2}{{q_n'}^2} \right) \right]
,\end{equation}
with $L_n$ being the total luminosity of the $n$-th component, $\sigma_n$ its dispersion along the $x'$ axis, and $q'_n$ its observed axial ratio. The sky co-ordinates $(x',y',z')$ were defined such that the $x'$ axis is aligned with the galaxy photometric major axis and the $z'$ axis coincides with the line of sight (LoS).

We fitted the MGE model to the HST photometry in F606W using the algorithm by \cite{cappellari02} and implemented in \texttt{MgeFit}\footnote{\url{https://pypi.org/project/mgefit/}}. In particular, we used the routine \texttt{mge\_fit\_sectors\_regularized}, which also implements the prescription from \cite{scott13} to avoid fitting bars and non-axisymmetric features. We used the routine \texttt{find\_galaxy} to get an initial guess for the semi-major axis position angle (PA), galaxy centroid, and ellipticity. The fit accounts for the point spread function of the image, which for the HST observations we assume to be a Gaussian with a deviation equal to 1 px, or equivalently a full width at half maximum of $\sim 2.4$ px. The fitted MGE coefficients are reported in Table~\ref{tab:1}.
The peak surface intensities, $I_n$, in units of [\!\lsol\ pc$^{-2}$] were obtained using the appropriate zero points for each observation\footnote{\url{https://acszeropoints.stsci.edu/}} and adopting an absolute magnitude for the Sun of $M_\odot=4.73$ mag \citep{2018ApJS..236...47W}. The intensities were corrected for Galactic extinction, and for the surface brightness dimming due to the cosmological redshift. For the Galactic extinction, we adopted $A_\mathrm{F606W}=0.307$ mag for NGC~4993 and $A_\mathrm{F606W}=0.035$ mag for NGC~1396, both computed from the Galactic dust reddening measured by \cite{sf17} assuming a visual extinction to reddening ratio of $R_V=3.1$ and the extinction law of \cite{fitzpatrick99}. 
To correct for the redshift dimming, we multiplied the intensities by $(1+z)^{3}$, which takes into account the $(1+z)^{4}$ factor from the bolometric dimming and the $(1+z)^{-1}$ factor, since the intensities are per unit frequency and not bolometric. 

The choice of co-ordinates made in Eq.~\ref{eq:1} implicitly assumes that all Gaussians have the same axes. The MGE formalism can be recast in a more general form in which the Gaussians have all different PAs with respect to each other \citep{monnet92,cappellari02}, but we do not find this necessary as our model already provides an acceptable fit. Under the assumption that the intrinsic shape is axisymmetric (which would not be allowed if different PAs were required for a good fit), the MGE model can be deprojected into the 3D stellar luminosity density with the simple analytical expression
\begin{equation}\label{eq:2}
\upsilon_\star(R,z)=\sum^{N}_{n=1}\frac{L_n}{\left(2\pi\sigma^2_n\right)^{3/2}q_n}\exp\left[ -\frac{1}{2\sigma_n^2} \left( R^2 +\frac{z^2}{q_n^2} \right) \right]
,\end{equation}
where $q_n$ are the intrinsic axial ratios,
\begin{equation}\label{eq:3}
q_n=\frac{\sqrt{{q'_n}^2-\cos^2 i}}{\sin i}
,\end{equation}
and $i$ is the galaxy inclination ($i=90^{\circ}$ being edge-on). The sky co-ordinates $(x',y',z')$ are related to the cylindrical co-ordinates $(R,z)$ by
\begin{equation}
\begin{pmatrix}
    x \\ y \\ z
\end{pmatrix}
=
\begin{pmatrix}
    1 & 0 & 0 \\
    0 & -\cos i & \sin i \\
    0 & \sin i & \cos i
\end{pmatrix}
\begin{pmatrix}
    x' \\ y' \\ z'
\end{pmatrix}
,\end{equation}
with $R^2=x^2+y^2$ and the $z$ axis being the symmetry axis of the galaxy. Although the axisymmetric MGE model provides a unique deprojection for a given inclination, one should bear in mind that the surface brightness deprojection is in general not unique \citep{1987IAUS..127..397R,1988MNRAS.231..285F}, and thus a specific choice of deprojection can introduce systematic errors in the inferred dynamical properties \citep[e.g.][]{2012MNRAS.424.1495L}.

\begin{table}
        \centering
        \caption{Coefficients of the MGE models fitted to NGC~4993 and NGC~1396.}
        \label{tab:1}
        \begin{tabular}{lccc}
                \hline\hline
                \noalign{\smallskip}
                $n$ & $\log I_n$ & $\log \sigma_n$ & $q'_n$\\
                & [\!\lsol\ pc$^{-2}$] & [arcsec] &\\   
                \noalign{\smallskip}
                \hline
                \noalign{\medskip}
                \multicolumn{4}{c}{NGC~4993}\\
                \noalign{\smallskip}
                1 & 4.261 & -1.173 & 0.894 \\
                2 & 3.677 & -0.362 & 1.000 \\
                3 & 3.426 & -0.039 & 0.880 \\
                4 & 3.069 & 0.373 & 0.850 \\ 
                5 & 2.674 & 0.713 & 0.850 \\
                6 & 2.089 & 1.068 & 0.850 \\
                7 & 1.354 & 1.340 & 1.000 \\
                8 & 1.107 & 1.537 & 1.000 \\
                \noalign{\smallskip}
                \hline
                \noalign{\medskip}
                \multicolumn{4}{c}{NGC~1396}\\
                \noalign{\smallskip}
                1 & 3.974 & -1.125 & 0.800 \\
                2 & 2.053 & -0.224 & 0.800 \\
                3 & 2.010 & 0.277 & 0.800 \\
                4 & 1.971 & 0.778 & 0.550 \\
                5 & 1.797 & 0.539 & 0.783 \\
        6 & 1.673 & 1.125 & 0.590\\
                \noalign{\smallskip}
                \hline
        \end{tabular}
\end{table}

\subsubsection{Stellar kinematics and galactic potential}

\begin{figure}
        \includegraphics[scale=1]{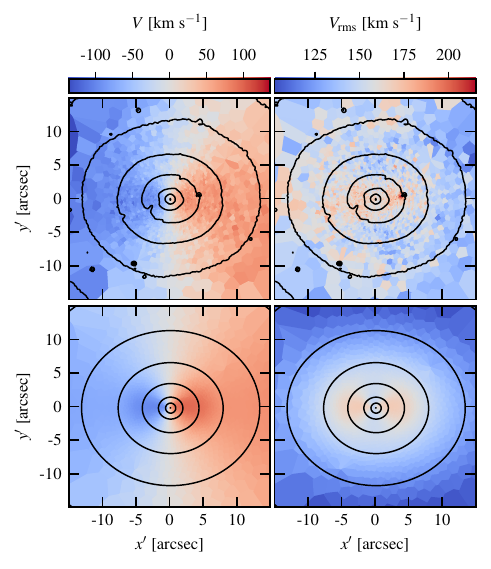}
    \caption{Projected kinematics of NGC~4993, inferred from MUSE data (top row) and predicted by JAM (bottom row). The left column shows the mean stellar velocity, $V$, along the LoS, while the right column shows the rms velocity, $V_\mathrm{rms}$, along the LoS. The solid black lines represent the observed isophotes (top row) and the MGE fit (bottom row). The $x'$ axis is aligned with the galaxy photometric major axis.}
    \label{fig:2}
\end{figure}

To model the stellar kinematics, we used the Jeans equations for the velocity distribution moments of a stellar system in steady state \citep{1922MNRAS..82..122J} through the JAM formalism \citep{cappellari08,cappellari20}.
We employed in particular the axisymmetric anisotropic Jeans equations derived by \cite{cappellari08} that describe the second-order moments of the velocities in cylindrical co-ordinates under the assumptions that the stellar system is axisymmetric, the velocity ellipsoid is aligned with the cylindrical frame, and the velocity anisotropy is constant.

Under these assumptions, the Jeans equations read
\begin{align}
\label{eq:5}\frac{b\upsilon\overline{v^2_z}-\upsilon\overline{v^2_\phi}}{R}+\frac{\partial(b\upsilon\overline{v^2_z})}{\partial R} &= -\upsilon\frac{\partial\Phi}{\partial R}\\
\label{eq:6}\frac{\partial(\upsilon\overline{v^2_z})}{\partial z} &= -\upsilon\frac{\partial\Phi}{\partial z}
,\end{align}
where $\Phi$ is the galactic gravitational potential, $\upsilon$ is the density distribution of the kinematic tracer, and $b$ parametrise the classical anisotropy parameter,
\begin{equation}
\beta_z\coloneqq 1-\frac{\overline{v^2_z}}{\overline{v^2_R}}=1-\frac{1}{b}
,\end{equation}
so that $\overline{v^2_R}=b\overline{v^2_z}$. Here, the second-order velocity moments are defined by
\begin{equation}
\upsilon\overline{v_i v_j}\coloneqq \int v_i v_j f\,\mathrm{d}^3\mathbf{v}
,\end{equation}
with $f$ being the distribution function such that $f(\mathbf{x},\mathbf{v})\,\mathrm{d}^3\mathbf{x}\,\mathrm{d}^3\mathbf{v}$ is the probability that a star is found in a given infinitesimal phase-space volume. With the boundary condition $\upsilon\overline{v^2_z}=0$ for $z\rightarrow \infty$, Eqs. \ref{eq:5} and \ref{eq:6} have solutions
\begin{align}
\upsilon\overline{v^2_R} &= b \upsilon\overline{v^2_z}\\
\upsilon\overline{v^2_\phi} &= b\left[ R\frac{\partial(\upsilon\overline{v^2_z})}{\partial R} +\upsilon\overline{v^2_z} \right] +R\upsilon\frac{\partial\Phi}{\partial R}\\
\upsilon\overline{v^2_z} &= \int^\infty_z\upsilon\frac{\partial\Phi}{\partial z}\,\mathrm{d}z
,\end{align}
which can be integrated along the LoS to get the projected second velocity moment, $\overline{v^2_\mathrm{los}}\coloneqq\overline{v^2_{z'}}$ 
\begin{equation}
\Sigma\overline{v^2_\mathrm{los}}=\int_{-\infty}^{+\infty} \left( \upsilon\overline{v^2_R}\sin^2 \phi + \upsilon\overline{v^2_\phi}\cos^2 \phi \right)\sin^2 i +\upsilon\overline{v^2_z}\cos^2 i \,\mathrm{d}z'
,\end{equation}
where $\cos\phi= x/R$. The second moment, $\overline{v^2_\mathrm{los}}$, approximates the observed root-mean-square (rms) velocity, $V_\mathrm{rms}\coloneqq \sqrt{V^2+\sigma^2}$, where $V$ and $\sigma$ are the observed mean stellar velocity and velocity dispersion, respectively, and thus it can be compared to the observed galactic kinematics to infer the free parameters of the JAM model.

To parametrise the tracer density, $\upsilon$, and the galactic potential, $\Phi$, we employed the MGE formalism, under the assumption that the kinematic and photometric axes coincide. 
As the tracer density, $\upsilon$, we used the stellar luminosity density, $\upsilon_\star$, which was deprojected from the MGE model of the stellar surface brightness (Eq.~\ref{eq:2}) and had the inclination, $i$, as a free parameter.
For the galactic potential, $\Phi$, we first parametrised the total density, $\rho$, as the sum of two 
sets of Gaussians,
\begin{equation}\label{eq:12}
\rho(R,z)= \left(\frac{M_\star}{L}\right)\, \upsilon_\star + \upsilon_\textsc{dm}
,\end{equation}
and then computed $\Phi$ from $\rho$ using the analytical formula for a density distribution constant on concentric homothetic ellipsoids \citep[§2.5.3]{bt} given by \citet[][see also Eqs.~16 and 17 of \citealt{cappellari08}]{emsellem94}. 
In the first term of Eq. \ref{eq:12}, we again have the stellar luminosity density, $\upsilon_\star$, this time multiplied by the stellar mass-to-light ratio, $M_\star/L$, to give the stellar mass density. The second term, on the other hand, is the mass density of a spherical dark matter (DM) halo, obtained by fitting a MGE model to the Navarro-Frenk-White (NFW) profile \citep{1997ApJ...490..493N},
\begin{equation}
\rho_\textsc{dm}(r)=\rho_0\, \left(\frac{r}{r_s}\right)^{-1}\left(1+\frac{r}{r_s}\right)^{-2}
,\end{equation}
with the routine \texttt{mge\_fit\_1d} from \texttt{MgeFit}, and deprojecting it as we did for the surface brightness with Eq. \ref{eq:2}. The NFW profile has two parameters -- namely, the normalisation factor, $\rho_0$, and the scale radius, $r_s$ -- however, only $\rho_0$ can be constrained with the stellar kinematics as the typical $r_s$ sits well outside the stellar light (see Sec.~\ref{sec:4.1}). 
Therefore, we fixed $r_s$ using scaling relations and used JAM to infer $\rho_0$, which we parametrised with the fraction of DM within the half-light radius,
$f_\textsc{dm}$.
The scale radius, $r_s$, was estimated by combining the stellar-to-halo mass relation of \cite{2013MNRAS.428.3121M} with the mass-concentration relation of \cite{2014MNRAS.441.3359D}. As an initial estimate, we used $31.0$ kpc for NGC~4993 and $7.8$ kpc for NGC~1396, computed assuming that the stellar mass of NGC~4993 is $M_\star=4.5\times 10^{10}\msol$ \citep{blanchard2017} and that of NGC~1396 is $M_\star=4.0\times 10^{8}\msol$ \citep{2016MNRAS.463.2819M}.

The JAM model described above was implemented by \cite{cappellari08} in \texttt{JamPy}\footnote{\url{https://pypi.org/project/jampy/}}. We used the routine \texttt{jam\_axi\_proj} to model $\overline{v^2_\mathrm{los}}$, which was then fitted to the $V_\mathrm{rms}$ inferred from the MUSE observations. The posterior distributions of $i,\,\beta_z,\,f_\textsc{dm}$, and $M_\star/L$ were inferred using the adaptive \cite{met} algorithm of \cite{haario}, as it is implemented by \cite{2013MNRAS.432.1709C} in \texttt{AdaMet}\footnote{\url{https://pypi.org/project/adamet/}}. The observed kinematics were computed from the MUSE IFS with \texttt{Starlight} \citep{2005MNRAS.358..363C}, following the method detailed in \cite{2018MNRAS.473.1359L}. As per the \texttt{Starlight} manual, we computed the velocity dispersion, $\sigma$, with $\sigma^2=\mathrm{vd}^2-\sigma^2_\mathrm{inst}+\sigma^2_\mathrm{base}$, where $\mathrm{vd}$ is the dispersion value inferred by \texttt{Starlight}, $\sigma_\mathrm{inst}$ is the spectral resolution of the observations, and $\sigma_\mathrm{base}$ is the resolution of the synthetic spectra. We assume that $\sigma_\mathrm{inst}=45$ km s$^{-1}$ as MUSE has a resolving power of $R\approx 2800$ at $7000$ \AA (corresponding to a resolution of 2.5 \AA), and $\sigma_\mathrm{base}=55$ km s$^{-1}$ as we fitted the spectra from \cite{2003MNRAS.344.1000B} that have a resolution of 3 \AA.

\begin{figure}
        \includegraphics[scale=1]{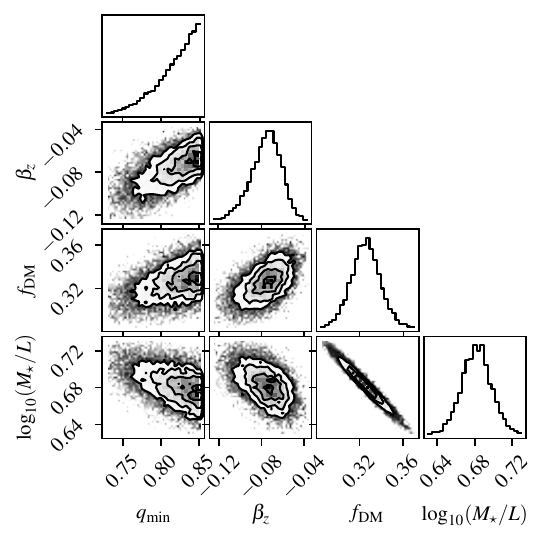}
    \caption{Posterior probability distributions of the JAM parameters inferred for NGC~4993 with \texttt{AdaMet}; namely, the minimum intrinsic axial ratio, $q_\mathrm{min}$, the anisotropy parameter, $\beta_z$, the fraction of DM within a half-light radius, $f_\mathrm{DM}$, and the logarithm of the stellar mass-to-light ratio, $\log_{10}(M_\star/L)$. The first parameter is defined as $q_\mathrm{min}=\min\left(\{q_n\}_{n=1,2,...,N}\right)$ and it parametrises the inclination, $i$, through Eq.~\ref{eq:3}. The distributions are represented marginalised over one and two dimensions (histograms and contours, respectively).
    The contours levels enclose the 12, 40, 68, and 86 percentiles (equivalent to the 0.5, 1, 1.5, and 2$\sigma$ envelopes in 2D).}
    \label{fig:2b}
\end{figure}

\subsection{Stellar population synthesis and supernova kicks}

The stellar populations that we have used in this work are from the Binary Population And Spectral Synthesis code \cite[\bpass;][]{eldridge17,stanway18, stevance20}, which uses a custom version of the Cambridge \texttt{STARS} code to perform the stellar evolution. 
We specifically employed the models of \cite{stevance2023}, characterised by a Kroupa initial mass function \citep{kroupa2001} and initial binary parameter distributions (mass ratios and periods) from \cite{moedistefano2017}. We refer the reader to \cite{eldridge2019} for an in-depth discussion on delay times and merger rates, and to \cite{stevance2023} for the properties of GW~170817 progenitors implied by the BPASS models.

As is described in \cite{stevance2023}, we can include the effects of supernova (SN) kicks a posteriori without recalculating a new grid of stellar models, using the \texttt{BPASS-TUI} code. The post-SN eccentricity, semi-major axis, and systemic velocity were computed using the formalism of \cite{tauris1998}, which takes into account the pre-SN orbital parameters, the mass ejected in the SN, and the natal kick. We assumed the natal kick directions to be isotropically distributed. The merger time was computed using the approximation of \cite{2021RNAAS...5..223M}.

A number of kick prescriptions have been used when modelling the effects of SN explosions, either based on observations of pulsar velocities \cite[e.g.][]{hobbs2005, verbunt2017, 2020MNRAS.494.3663I} or theoretically motivated by the physics of the explosion \cite[e.g.][]{bray2016,bray2018,2020ApJ...891..141G,mandel2020}. In the case of the present work, it would be computationally prohibitive to attempt to perform our analysis using all the kick prescriptions that have been proposed in the literature. Indeed, our synthetic population counts 446444 BNS merger precursors that must then each be evolved through galactic potentials with randomised initial positions to recover final offsets.

Consequently, we have limited our choice of kick prescription to two ‘extreme’ cases that allow us to explore the effects of kicks on our final offsets.
Firstly, we used the single-peaked Maxwellian from \citet{hobbs2005}, which is known to be an overestimate. 
In follow-up models, we then set all the kicks after the second SN to have $v_\mathrm{ kick}=30$\kms, which is about the kick velocity typically used when considering electron-capture or ultra-stripped SNe \citep[e.g.][]{2004ApJ...612.1044P,2018MNRAS.481.1908K,vigna-gomez2018,2018MNRAS.480.2011G}. 
Although these small kicks are consistent with the properties of observed Galactic BNSs \citep[e.g.][and references therein]{2016MNRAS.456.4089B,2017ApJ...846..170T}, this is not a physical solution as the vast majority of SNe in these models do not fit electron-capture or ultra-stripped conditions, but it serves as a deliberate underestimate to see how this affects the final offsets. 

\begin{figure}
    \centering
    \includegraphics[scale=1]{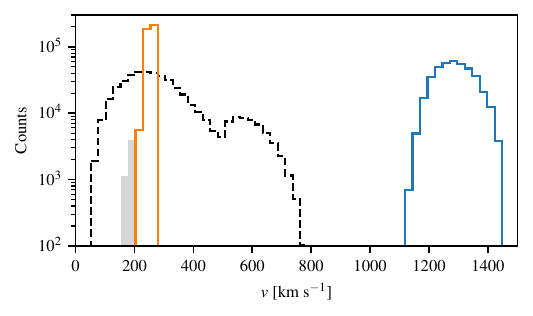}
    \caption{Systemic kicks from SN2 and escape velocities in our models. The dashed black line is the systemic kicks distribution predicted by \texttt{BPASS-TUI}. The shaded grey areas are the systems that received a natal kick $v_\mathrm{k}$ of 30 km s$^{-1}$. The solid coloured lines are the escape velocities at the locations of SN2, for both galaxies.}
    \label{fig:5}
\end{figure}

In Fig.~\ref{fig:5}, we show a comparison between the systemic kicks from the second SN, and the escape velocities at the location of the second SN. For reference, the circular velocity at these locations is on average $\sim4-5$ times lower than the escape velocity.

\subsection{Galactic trajectories and merger locations of binary neutron stars}

To model the BNS merger locations, we seeded the BPASS population in the galactic potential inferred with JAM, and simulated the BNS trajectories with \texttt{galpy}\footnote{\url{https://github.com/jobovy/galpy}} \citep{bovy15}. The potential was built summing the NFW potential with the stellar potential, the former obtained using the routine \texttt{potential.NFWPotential} and the latter obtained approximating the MGE model (i.e. the first term in Eq. \ref{eq:12}) with the self-consistent field method of \cite{hernquist92} using the routine \texttt{potential.SCFPotential}. We assume that the potential remains constant over time, which is a reasonable assumption at least for the stellar component of NGC~4993, since its star formation ceased around 9 Gyr ago \citep{blanchard2017,levan2017,stevance2023}.

To seed the binaries, we computed the stellar mass enclosed in spherical shells by integrating radially Eq. \ref{eq:2} times $M_\star/L$, and then seeded the binaries at random locations using the enclosed mass as a proxy. The binaries were initialised on circular orbits and immediately kicked with the systemic kicks from the first SN (hereafter SN1) given by \texttt{BPASS-TUI}. The galactic trajectories were integrated up to the time of the second SN (hereafter SN2) when the binaries were kicked again, and finally integrated up to the GW-driven merger to get the merger locations. We simulated one trajectory per BNS model for a total of 446444 trajectories, randomly sampling for each binary a different initial location and a different orientation for each systemic kick.


\begin{figure}
    \centering
    \includegraphics[scale=1]{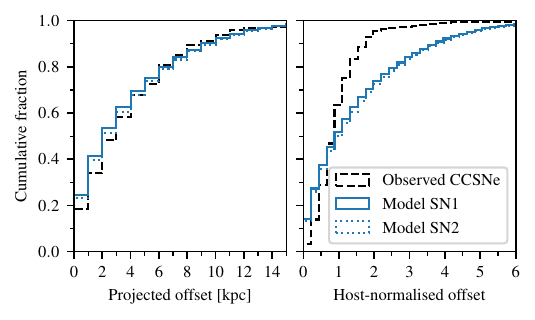}
    \caption{Comparison between observed and modelled locations of core-collapse supernovae (CCSNe). \emph{Left.} CCSN offsets from the ATLAS volumetric survey (\citealt{srivastav22}; Srivastav et al. in prep) compared to those seen in our models (note that these are exclusively binary systems that can lead to BNS mergers, not an exhaustive theoretical population of CCSNe). \emph{Right.} Same as in the left panel, with the observed normalised offsets of CCSNe from \citet{2012ApJ...759..107K}.}
    \label{fig:3}
\end{figure}

\begin{figure*}
    \centering
    \includegraphics[scale=1]{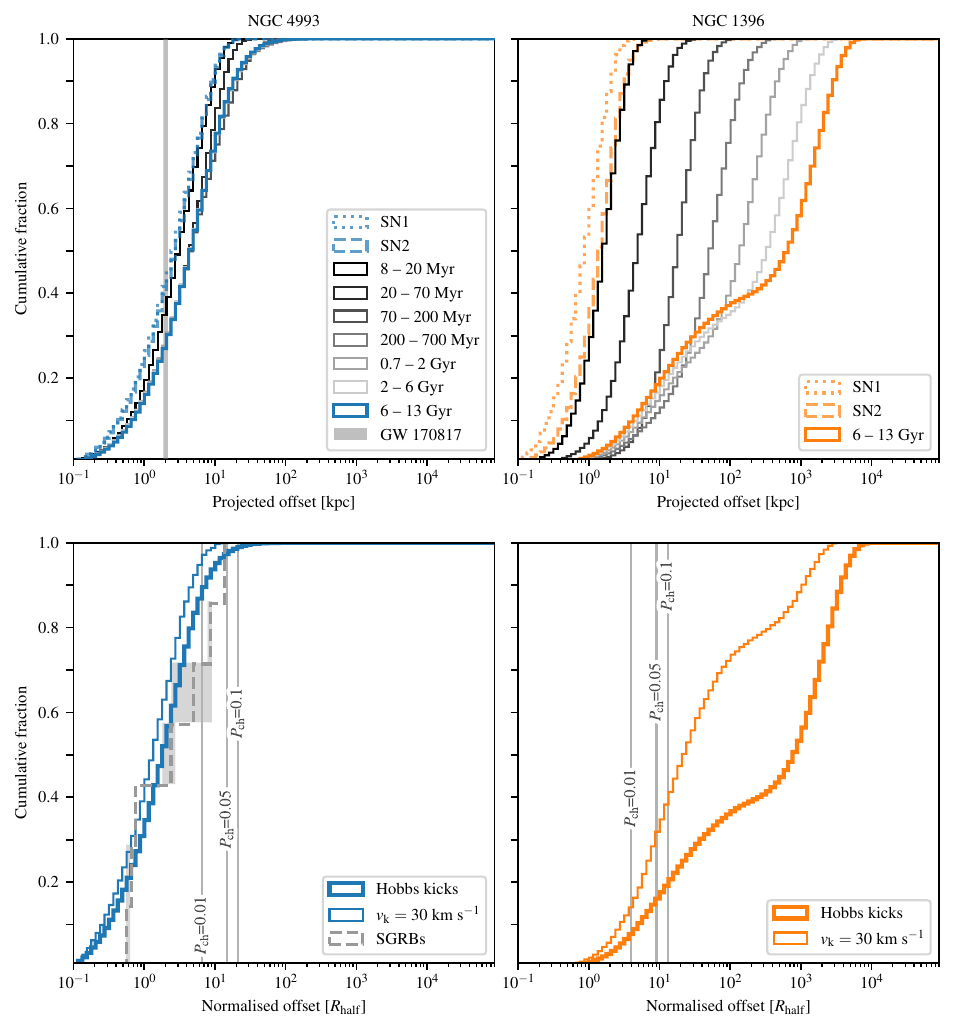}
    \caption{Location of BNS mergers predicted by our models. \emph{Top.} Projected offsets of BNS mergers in physical units. The dashed and dotted lines indicate the locations of SN1 and SN2, respectively, while the remaining lines represent the BNS merger offsets divided in log-spaced time bins. The shaded vertical band indicates the observed projected offset of GW~170817. \emph{Bottom.} Host-normalised offsets of BNS mergers with delay times between 5 and 13 Gyr (thick solid lines). This time bin matches the age of possible GW~170817 progenitors. The thin lines show the distribution of merger offsets produced by BNSs where the neutron stars received a small natal kick, $v_\mathrm{k}$, of 30 km s$^{-1}$ at SN2. The dashed line represents the distribution of normalised offsets for the SGRBs in the BRIGHT catalogue \citep{fong2022,nugent2022} with a host galaxy more massive than NGC~4993 (the shaded region indicates the offsets' uncertainties). The vertical lines labelled with $P_\mathrm{ch}$ identify the normalised offsets at which the probability of chance alignment, $P_\mathrm{ch}$ \citep{bloom2002}, with each galaxy is either 1, 5, or 10 percent.}
    \label{fig:4}
\end{figure*}

\section{Results}\label{sec:4}

\subsection{NGC~4993 properties inferred from stellar kinematics}\label{sec:4.1}

The MGE fit to the surface brightness of NGC~4993 is reported in Table~\ref{tab:1}. 
The fit has an rms error of $\sim 3$ percent, and has been corrected for Galactic extinction and redshift dimming. The MGE total luminosity is $L_{606} = 1.43 \times 10^{10}\lsol$, equivalent to an absolute magnitude of $M_{606} = -20.69$, while the half-light radius, $r_\mathrm{half}$, is $13.51$ arcsec, or $2.64$ kpc at the galaxy redshift. 

The JAM model is shown in Fig.~\ref{fig:2}, while the posterior distributions of its parameters are shown in Fig.~\ref{fig:2b}. From the kinematics, we infer an inclination, $i=88.58^{+1.42}_{-20.22}$ deg, an anisotropy parameter, $\beta_z=-0.07\pm0.02$, a fraction of DM, $f_\textsc{dm}=0.33\pm0.01$, and a stellar mass-to-light ratio of $M_\star/L=4.74\pm0.18\,\msol\lsol^{-1}$. From the total luminosity and the mass-to-light ratio, we infer that NGC~4993 has a stellar mass of $M_\star=6.76^{+0.46}_{-0.43}\times10^{10}\,\msol$. 

For the NFW profile, we inferred the scale radius, $r_s=46.10^{+22.01}_{-15.43}$ kpc, from the stellar mass, and the normalisation factor, $\rho_0=7.16^{+2.68}_{-1.79}\times10^{-3}\,\msol\,\textrm{pc}^{-3}$, from $M_\mathrm{half}$ and $f_\textsc{dm}$. With these parameters, the DM halo has a mass of $M_{200}=10.22^{+13.18}_{-5.85}\times 10^{12}\,\msol$.
All the errors quoted above were obtained propagating the uncertainties on the JAM parameters and the MGE rms error. We have not propagated the uncertainty on the redshift, since its contribution is negligible.

We compared these dynamical estimates to the literature values. \cite{abbott2017_host} uses $M_\star=6.21\times10^{10}\,\msol$ and $M_\mathrm{vir}=3.46\times10^{12}\,\msol$, both of which were obtained from scaling relations. \cite{im2017} uses $M_\star=3-12\times10^{10}\,\msol$, \cite{palmese2017} uses $M_\star=3.8\pm0.2\times10^{10}\,\msol$, \cite{blanchard2017} uses $M_\star=4.47^{+0.32}_{0.30}\times10^{10}\,\msol$, and \cite{levan2017} uses $M_\star=14\times10^{10}\,\msol$, all of which were obtained from fitting the spectra or the spectral energy distribution. \cite{ebrova2020} uses $M_\mathrm{vir}=1.9^{+1.2}_{-0.7}\times10^{12}\,\msol$, from scaling relations. \cite{palmese2017} also gives $M_\star/L=5.23\pm0.15$ in the $r$ band. Overall, our stellar mass estimate sits well within the spread of mass estimates from the literature, while our DM halo mass is more than $1\sigma$ greater than the estimates of \cite{abbott2017_host} and \cite{ebrova2020}. We note that a difference in the dynamical mass of a factor, $f$, implies a change in the galaxy circular and escape velocities by a factor of $\sim\sqrt{f}$. 

Since we have integral-field kinematics, we also computed the luminosity-weighted estimate for the projected specific angular momentum of \cite{emsellem+07}
\begin{equation}
\lambda_R=\frac{\langle R |V| \rangle}{\langle R\sqrt{V^2+\sigma^2} \rangle}=\frac{\sum_{n=1}^{N}F_nR_n|V_n|}{\sum_{n=1}^{N}F_nR_n\sqrt{V^2_n+\sigma^2_n}}
\end{equation}
and the luminosity-weighted projected ellipticity,
\begin{equation}
\varepsilon^2 = 1-\frac{\langle y^2 \rangle}{\langle x^2 \rangle} = 1-\frac{\sum_{n=1}^{N}F_ny^2_n}{\sum_{n=1}^{N}F_nx^2_n}
,\end{equation}
where $F_n,\,R_n,\,V_n,\,\sigma_n,x_n,y_n$ are, respectively, the bolometric flux, projected radius, mean stellar velocity, velocity dispersion, and Cartesian co-ordinates along the photometric axes of the $n$-th Voronoi bin. The values within a $1R_\mathrm{half}$ aperture for NGC~4993 are $\lambda_{R}(R_\mathrm{half})=0.35$ and $\varepsilon(R_\mathrm{half})=0.07$; thus, the galaxy is a relatively round fast rotator, which is consistent with it being an S0 seen almost face-on \citep{emsellem+11,cappellari16,2018MNRAS.477.4711G}. This value of $\lambda_R$, however, is relatively high compared to that of other early-type galaxies (ETGs) with shell systems. \cite{yoon2024} analysed a sample of nearby ETGs from the MaNGA survey, finding that half of ETGs with shells are classified as slow rotators (i.e. have $\lambda_R \lesssim 0.1-0.15$). This fraction is three times the fraction of slow rotators in ETGs with no tidal features, and twice the fraction in ETGs with other tidal features, such as tails and streams. We also note that ETGs with dust lanes have higher $\lambda_R$ than ETGs without dust lanes \citep{yoon2022}, and that NGC~4993 shows clear although not prominent dust lanes in its very inner regions.

\subsection{Locations of supernovae and binary neutron star mergers in NGC~4993}\label{sec:4.2}

To report and analyse the location of transients in and around their host galaxies, we made use of two observables: the projected offset and the host-normalised offset. Defining the spherical offsets as the spherical radius of the transient location in the galactocentric frame, the projected offset is then the spherical offset projected on the sky plane, while the normalised offset is the projected offset normalised by the projected half-light radius of the host galaxy.

In Fig.~\ref{fig:3}, we show the projected and normalised offsets of all the SNe in NGC~4993 predicted by our model. Since in our models SN1 occurs while the binary is still on the initial circular orbit, the locations of SN1 coincide with the light radial profile except for projection effects. The offset distributions of SN1 and SN2 almost overlap each other, indicating that the systemic kicks from SN1 are not strong enough to dislodge the binary from the SN1 location by the time SN2 happens. In Fig.~\ref{fig:3}, we also show the locations of observed core-collapse SNe for comparison, using the projected offsets from the ATLAS survey (volumetrically complete to 100 Mpc, Srivastav et al. in prep.) and the normalised offsets of \citet{2012ApJ...759..107K}. We emphasise that here we are not looking for a perfect match to observations: first of all, our simulations only look at a sub-sample of SN progenitors; secondly, we are looking at the effects of a specific galactic potential rather than a range of galaxies like in the observations. We indeed find a good match between models and observations for projected offsets, although the match is not as good for normalised offsets where we account for the size and light distribution of the host galaxy.

In the left panel of Fig.~\ref{fig:4}, we show the projected offsets of BNS merger in NGC~4993 predicted by our model. The fraction of BNS mergers predicted within the observed offset of GW~170817 is 24.4$^{+1.5}_{-1.6}$ percent. The mergers with a delay time compatible with that of GW~170817, namely between 6 and 13 Gyr \citep{levan2017,blanchard2017,stevance2023}, have a median projected offset of 4.8$^{+9.9}_{-3.7}$ kpc (here we are reporting the 16th-84th percentiles as uncertainties), which is compatible within 1$\sigma$ with the observed offset. This is only twice the offset at which they formed, as the median projected offset of SN1 is 2.7$^{+5.3}_{-2.0}$ kpc and the median normalised offset of BNS merger is indeed 1.9$^{+3.8}_{-1.4}$ (SN1 traces stellar light in our models). 

We note that the offset distribution does not evolve with respect to the delay times, as was also found by \cite{abbott2017_host} (see the top panels of their Fig.~5). The absence of time evolution together with the small displacement between formation and merger reflects the fact that all BNSs remain bound to the host (see Fig.~\ref{fig:5} for a comparison between systemic kicks and escape velocities). 

\begin{figure*}
    \centering
        \includegraphics[scale=1]{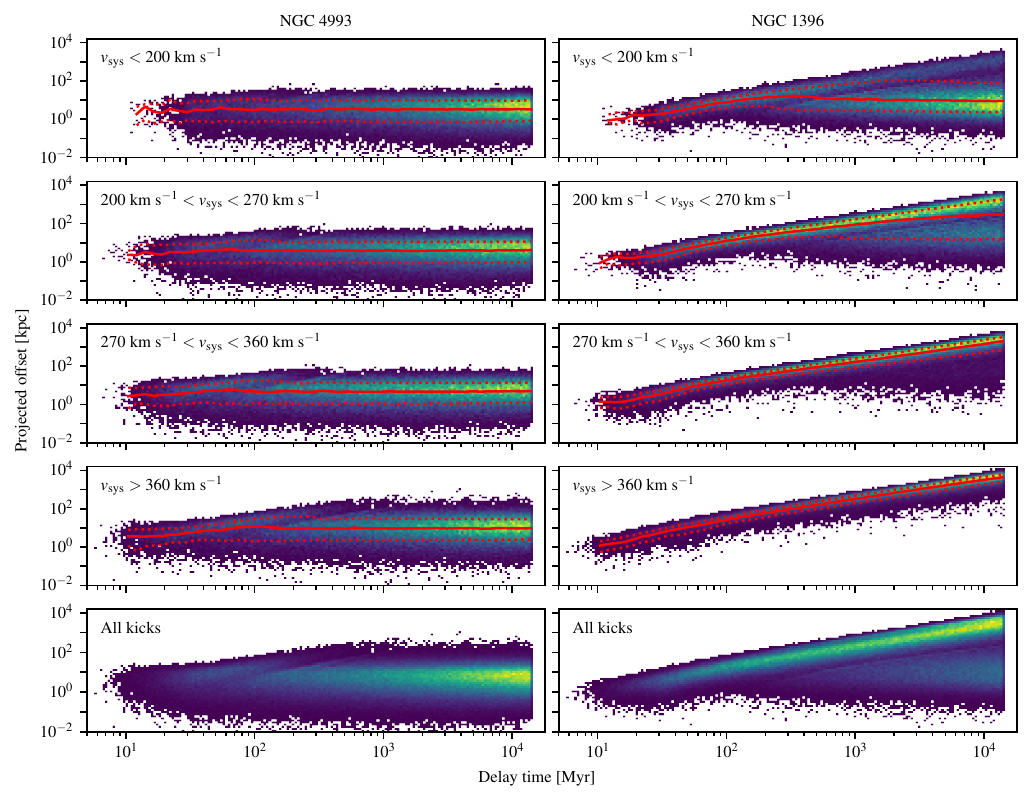}
    \caption{Histograms of the BNS mergers in the projected offset vs delay time plane. The systemic kicks from SN2 have approximately 200, 270, and 360 km s$^{-1}$ as 1st, 2nd, and 3rd quartiles; thus, each panel in the first four rows contains around 25\% of the mergers. The last row shows the whole sample. The two columns separate the results for NGC~4993 and for NGC~1396. The histograms are not normalised; hence, they show counts and not densities. The solid red lines represent the median offset, while the dotted red lines indicate the 16th-84th percentiles, both as a function of the delay time.}
    \label{fig:6}
\end{figure*}

\subsection{Comparison to the locations of BNS mergers in the potential of a dwarf galaxy}\label{sec:4.3}

To understand the impact of the potential on the predicted merger offsets, we repeated the simulation, swapping the potential and light distribution of NGC~4993 with that of the dwarf galaxy NGC~1396. In this simulation, we seeded the BNSs, following the light distribution of NGC~1396 and using its circular velocity. The merger offsets predicted for NGC~1396 are shown in the central panel of Fig.~\ref{fig:4}.

For the dwarf galaxy, we find that 37\% of all BNS remain bound to the galaxy after the kicks, unlike for NGC~4993 where all systems remained bound. When considering only the BNS mergers compatible with the delay time of GW~170817, the bound BNSs account for 42\% of the mergers and merge at a median projected offset of 14.9$^{+374.9}_{-11.9}$ kpc, or at a median normalised offset of 18.6$^{+468.7}_{-14.8}$ (reporting again the 16th-84th percentiles as uncertainties). The unbound systems instead merge at a median projected offset of 1777.4$^{+1813.6}_{-1248.9}$ kpc, or at a median normalised offset of 2221.7$^{+2267.0}_{-1564.1}$. Both distributions of offsets evolve with time (see Fig.~\ref{fig:4}), with the bound systems eventually settling on a certain distribution, and the merger offsets of unbound increasing monotonically with time.

In our models, the systemic kick from SN1 is $78^{+52}_{-33}$ km s$^{-1}$ (median value with 16th-84th percentiles as uncertainties), which is negligible compared to the escape velocity of NGC~4993 and only a fraction of the escape velocity of NGC~1396 (see Fig.~\ref{fig:5}). Therefore, we expect virtually all BNSs progenitors to remain bound to the host after SN1. When multiplying the kicks by the time between SN1 and SN2, we obtain a distance of $1.5^{+1.0}_{-0.6}$ kpc (median value with 16th-84th percentiles as uncertainties). These distances are larger than the offset between SN1 and SN2 in NGC~4993 (see Fig.~\ref{fig:4}), and comparable but still greater than those in NGC~1396. This again points towards the fact the BNSs do not get unbound upon SN1, and hence that we cannot estimate the offset between SN1 and SN2 by simply multiplying velocity and time of flight, although we notice that in a shallower potential the orbits are much more perturbed.

In the right panel of Fig.~\ref{fig:4}, we compare the prediction for NGC~4993 to those for the dwarf galaxy. 
The predicted BNS mergers in NGC~4993 have normalised offsets that are comparable to those of the observed SGRBs with hosts of a stellar mass greater than or equal to NGC~4993 in the BRIGHT catalogue \citep{fong2022,nugent2022}. In the dwarf, on the other hand, bound systems merge at normalised offsets that are $\sim 1$ dex greater than in NGC~4993, and up to $\sim 3$ dex when considering unbound systems. For such high offsets, the probability of chance alignment, $P_\mathrm{ch}$ \citep{bloom2002}, of the dwarf is always 1 (see Fig.~\ref{fig:4}, assuming both the galaxy and the merger electro-magnetic counterpart are detectable), and therefore it poses a bias against identifying the dwarf as the host. This would not be true for bright hosts like NGC~4993, since the association by $P_\mathrm{ch}$ is more robust even for the upper tail of their offset distribution \citep[][also Fig.~\ref{fig:4}]{berger2010, tunnicliffe2014}, 
whereas the $P_\mathrm{ch}$ of faint hosts would be small only if both the offset and its error region were small \citep[e.g.][]{levan2007, 2017ApJ...849..162E}.
In Fig.~\ref{fig:4}, we show for reference the $P_\mathrm{ch}$ from \cite{bloom2002} for both galaxies, computed assuming an apparent magnitude of $R=11.9$ for NGC~4993 \citep{1989spce.book.....L} and $R=14.2$ for NGC~1396 (obtained from the $g'r'i'$ magnitudes of \citealt{2018AA...620A.165V} using the photometric transformations of \citealt{2005AJ....130..873J}). We also note that the fraction of bound to unbound systems depends on both the systemic kicks and the choice of potential, as is shown in the right panel of Fig.~\ref{fig:4}.

\subsection{Merger offsets versus systemic kicks and delay times}\label{sec:4.4}

In Fig.~\ref{fig:6}, we show the distribution of BNS mergers on the offset versus delay time plane, as a function of the potential and the systemic kick from SN2. For NGC~4993 (left panels in Fig.~\ref{fig:6}), the mergers are distributed on a horizontal strip in the plane, with the 16-50-84 percentiles of offsets being constant across all delay times beyond $\sim 100$ Myr. This reflects the lack of time evolution in the offset distribution (see Sec.~\ref{sec:4.2}). The offset range, however, shows a dependence on the systemic kick, with the offset distribution shifting toward larger values for increasing kick bins (Fig.~\ref{fig:6}). Therefore, while the small offset of GW~170817 provided a robust host association via $P_\mathrm{ch}$, it also leaves unconstrained both the systemic kick and the delay time, as it sits within the bulk population independently of the range of systemic kicks or delay times. The largest offsets instead can only be reached with the highest kicks (see Fig.~\ref{fig:6}) and can therefore be used to probe the upper tail of the kick distribution, for instance through the energy argument of \cite{bloom2002}, since these merging systems are bound.
Previous works have already noted that the small offset of GW~170817 does not provide a strong constraint on the systemic kick \citep{blanchard2017,levan2017,pan2017,andreoni2017}. This is clearly shown in Fig.~5 of \cite{abbott2017_host}, where we see that the inferred posteriors are similar to the priors from the synthetic population given the observed offset (middle panels), and that they would start to probe higher natal kicks and progenitor masses only if the offset gets close to 100 kpc (bottom panels). A similar result is found by \cite{zevin2020} when modelling the large offsets of GRB~070809 and 090515, namely that large merger offsets can indeed push the constraint on the systemic kick towards higher values, and thus probe specific progenitors (see their Fig.~5 and 6).

In the case of the dwarf galaxy (right panels in Fig.~\ref{fig:6}), the BNS mergers split into two branches: one made by bound systems and another by unbound systems. The mergers of bound systems behave like those in NGC~4993; namely, forming a horizontal strip in the plane with little to no time evolution beyond a few 100 million years. The mergers of unbound systems instead form a diagonal strip, with a slope around unity for the highest kicks that suggests a linear relation between offsets and delay times. In Fig.~\ref{fig:6}, we see that the horizontal branch dominates for systemic kicks below 200 km $^{-1}$, while it gradually gives way to the diagonal sequence as the kick magnitude increases.

In the dwarf galaxy, the merger offsets of unbound systems quickly become too large to secure a host association by $P_\mathrm{ch}$ (see right panel of Fig.~\ref{fig:4}), while those of bound systems can remain small enough to secure an association (see e.g. \citealt{2024ApJ...962....5N} for a sample of dwarf hosts associated with $P_\mathrm{ch}$). For these smaller offsets, we can distinguish two extreme situations. 
If the delay time is short, we find BNSs merging at small offsets regardless of the systemic kick, and thus the offset alone is not effective at constraining the kick (see Fig.~\ref{fig:6}, e.g. at delay times below a few 100 million years). This might be the case in a young or star-forming host, where we expect to observe mergers with the shortest delay times, and therefore where it might be difficult to distinguish whether a small merger offset is due to a systemic kick below the escape velocity, or a short delay time.
However, if the delay time is large then a small merger offset is only compatible with a bound system, since mergers of unbound systems would be too far out for association. This is shown in Fig.~\ref{fig:6}, where at delay times above, for example, a few 100 million years, we find BNSs merging at small offsets only if the systemic kicks are about or below the escape velocity. Therefore, if the host is an old or quiescent galaxy, where we expect to be probing the upper tail of the delay time distribution, then a small merger offset could provide a systemic kick estimate given the typical escape velocity (that should account also for the progenitors' locations and velocities within the potential).

A caveat of our method is that we evolve BNS trajectories in a static galactic potential. Galaxies are known to experience mass and size growth over cosmic time \citep[e.g.][]{vanderwel2014}, and although the role of galaxy mergers in the growth of the stellar components seems dominant only for the most massive galaxies \citep[e.g.][]{rodriguezgomez2015,rodriguezgomez2016}, the effects of galaxy evolution on the galaxy structure and kinematics is still an open topic of research \citep{conselice2014,cappellari16}. 
A few works have already investigated the effects on the merger distribution of compact object binaries.
\cite{zemp2009} found that in an environment with a higher galaxy density such as a cluster, compact binaries do not reach offsets as high as in an isolated field galaxy due to the deeper potential, although the mixing between neighboring populations leads to a greater chance of misidentifying the true host. This effect is enhanced by large kicks \citep{kelley2010} and is pronounced for compact objects mergers happening in the local Universe \citep{wiggins2018}, while mergers at $z=2-3.5$ still trace the progenitor population.


\section{Conclusions}\label{sec:5}

We have discussed in detail the offset of GW~170817 with respect to its host galaxy NGC~4993, and the constraints this puts on the kinematics of merging BNSs. 
The stellar kinematics of NGC~4993 was modelled from IFS with JAM \citep[JAM,][]{cappellari08} in order to constrain the galactic potential. 
A synthetic population of BNSs was simulated with \texttt{BPASS-TUI} \citep{stevance2023} and seeded within NGC~4993's potential. The BNS galactic trajectories were computed with \texttt{galpy} \citep{bovy15} and their merger offsets were studied. We repeated the orbit simulation after switching the potential of NGC~4993 with that of a dwarf galaxy, to understand the different impacts of galactic potentials and systemic kicks in shaping the distribution of BNS merger offsets. We summarise our conclusions:

\begin{enumerate}
\item We infer that NGC~4993 has a dynamical stellar mass of $\sim 6.8\times10^{10}\,\msol$, a stellar mass-to-light ratio of $\sim 4.7\,\msol\lsol^{-1}$, and a DM halo mass of $\sim 10.2\times 10^{12}\,\msol$, which is consistent with previous estimates obtained from the galaxy spectrum and magnitudes \citep[e.g.][and references therein]{abbott2017_host,palmese2017,ebrova2020}. From the kinematics, we also determined that the galaxy is a quite round fast rotator, likely seen edge-on. We provide a model for the surface brightness and the galactic potential in Sec.~\ref{sec:4.1}.
\item Virtually all of the BNSs in our model remain bound to NGC~4993 after receiving two systemics kicks, even assuming that all natal kicks are drawn from the \cite{hobbs2005} distribution. The offset distribution of BNS mergers in NGC~4993 shows no time evolution and only a modest displacement between the first SN and the merger (see Fig.~\ref{fig:4}, left panel). The merger offset distribution is consistent with the locations of SGRBs in hosts with a stellar mass equal to or greater than NGC~4993 (see Fig.~\ref{fig:4}, right panel).
\item The observed offset of GW~170817 is consistent with our predictions regardless of large or small kicks, because the strong potential is only diagnostic of very large kicks (see Fig.~\ref{fig:4} and ~\ref{fig:6}).
Small offsets are not very constraining, whereas the largest can only be reached with the highest kicks, restricting the range of progenitors \cite[as seen by][]{abbott2017_host,zevin2020}.
\item For comparison, we evolved the BNS trajectories in the dwarf galaxy NCG~1396. Only around half of the systems remain bound to this host, merging at offsets $\sim1$ dex larger than those in NGC~4993. Of the bound systems, fewer than half of mergers result in a probability of chance alignment of $P_\mathrm{ch}<0.05$. The remaining unbound systems merge at offsets that correlate with the delay time and have $P_\mathrm{ch}\sim1$ (Fig.~\ref{fig:4}, right panel). We find that for delay times above $\sim1$ Gyr, the small merger offsets of bound BNSs can still lead to association by $P_\textrm{ch}$, and would effectively constrain the second kick to be below the escape velocity at the time of the SN. We find the fraction of bound systems to be determined by the kick distribution.
\end{enumerate}

The method presented in this paper is complementary to the end-to-end analysis of \cite{stevance2023}, which gives priors for the progenitor BNS based on the IFS of the host. A possible follow-up work is its application to a sample of SGRBs.


\begin{acknowledgements}
We thank the anonymous referee for the constructive comments.
We are grateful to the ATLAS team for early access of the Transient Volumetric Survey. 
We thank Michele Cappellari for the assistance given in the use of \texttt{jampy}.
NG acknowledges studentship support from the Dutch Research Council (NWO) under the project number 680.92.18.02.
HFS is supported by the Eric and Wendy Schmidt AI in Science Fellowship. 
AJL was supported by the European Research Council (ERC) under the European Union’s Horizon 2020 research and innovation programme (grant agreement No.~725246). 
AAC acknowledges support from the European Space Agency (ESA) as an ESA Research Fellow.
JDL acknowledges support from a UK Research and Innovation Fellowship (MR/T020784/1).
\end{acknowledgements}

\bibliographystyle{aa}
\bibliography{biblio}


\begin{appendix}
\section{NGC~1396 properties inferred from stellar kinematics}\label{sec:A.1}

The MGE fit to the surface brightness of NGC~1396 is reported in Table~\ref{tab:1}. 
The fit has a rms error of $\sim 11$ percent, and is corrected for Galactic extinction and redshift dimming. The MGE total luminosity is $L_{606} = 4.50 \times 10^{8}\lsol$, equivalent to an absolute magnitude $M_{606} = -16.90$, while the half-light radius $r_\mathrm{half}$ is $8.37$ arcsec, or $0.80$ kpc at the galaxy redshift. 

\begin{figure}
        \includegraphics[scale=1]{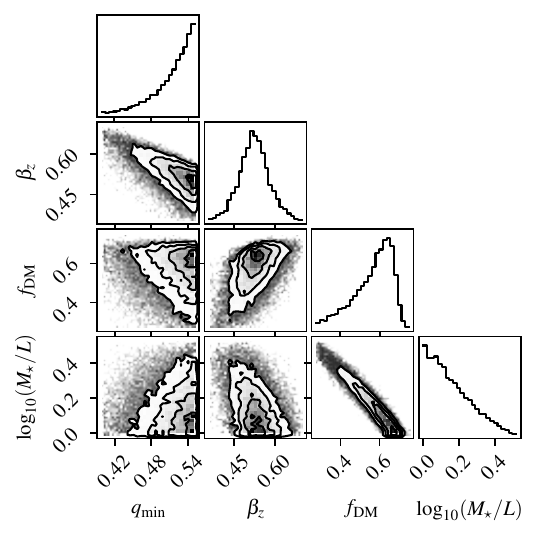}
    \caption{Posterior probability distributions of the JAM parameters inferred for NGC~1396 with \texttt{AdaMet}, namely the minimum intrinsic axial ratio $q_\mathrm{min}$, the anisotropy parameter $\beta_z$, the fraction of DM within a half-light radius $f_\mathrm{DM}$, and the logarithm of the stellar mass-to-light ratio $\log_{10}(M_\star/L)$.
    The contours levels enclose the 12, 40, 68 and 86 percentiles (equivalent to the 0.5, 1, 1.5, and 2$\sigma$ envelopes in 2D).}
    \label{fig:A1}
\end{figure}

The posterior distributions of the JAM parameters are shown in Fig.~\ref{fig:A1}. From the kinematics we infer an inclination $i=86.79^{+3.21}_{-9.45}$ deg, an anisotropy parameter $\beta_z=0.52\pm0.05$, a fraction of DM $f_\textsc{dm}=0.67\pm0.10$, and a stellar mass-to-light ratio $M_\star/L=1.03^{+0.33}_{-0.25}\,\msol\lsol^{-1}$. From the total luminosity and the mass-to-light ratio we infer that NGC~1396 has a dynamical mass $M_\star=4.65^{+2.13}_{-1.51}\times10^{8}\,\msol$. 

For the NFW profile, we infer the scale radius $r_s=8.02^{+2.03}_{-1.58}$ kpc from the dynamical mass, and the normalisation factor $\rho_0=10.23^{+8.46}_{-4.32}\times10^{-3}\,\msol\,\textrm{pc}^{-3}$ from $M_\mathrm{half}$ and $f_\textsc{dm}$. With these parameters, the DM halo has a mass $M_{200}=10.15^{+23.95}_{-6.92}\times 10^{10}\,\msol$.
All the errors quoted above are obtained propagating the uncertainties on the JAM parameters and the MGE rms error. We do not propagate the uncertainty on the redshift since its contribution is negligible.

\end{appendix}

\end{document}